\documentclass[preprint,12pt]{elsarticle}

\biboptions{numbers,sort&compress}

\usepackage{multicol}

\usepackage{lipsum}

\usepackage{bbm}

\usepackage{amssymb}

\usepackage[T1]{fontenc}

\usepackage{amssymb}

\usepackage{amsmath}

\usepackage[scientific-notation=false]{siunitx}

\counterwithin{figure}{section}
\counterwithin{table}{section}
\counterwithin{equation}{section}

\usepackage{setspace}

\usepackage{booktabs}

\usepackage{array}
\newcommand{\PreserveBackslash}[1]{\let\temp=\\#1\let\\=\temp}
\newcolumntype{C}[1]{>{\PreserveBackslash\centering}p{#1}}
\newcolumntype{R}[1]{>{\PreserveBackslash\raggedleft}p{#1}}
\newcolumntype{L}[1]{>{\PreserveBackslash\raggedright}p{#1}}
\usepackage{arydshln}

\journal{arXiv}

\begin{document}

\begin{frontmatter}



\title{Essential metrics for Life on graphs}


\author[ugent]{Michiel Rollier}
\ead{michiel.rollier@ugent.be}

\author[icmc]{Lucas Caldeira de Oliveira}

\author[ifsc]{Odemir M.~Bruno}

\author[ugent]{Jan M.~Baetens}

\affiliation[ugent]{
  organization={BionamiX,
  Dept.~of Data Analysis and Mathematical Modelling,
  Ghent University},
  addressline={Coupure Links, 653},
  city={Ghent},
  postcode={9000},
  country={Belgium}
}

\affiliation[icmc]{
  organization={Institute of Mathematical and Computer Sciences, University of S\~ao Paulo},
  addressline={Av.~Trab.~S\~ao Carlense, 400},
  city={S\~ao Carlos},
  postcode={13566-590},
  country={Brazil}
} 

\affiliation[ifsc]{
  organization={S\~ao Carlos Institute of Physics, University of S\~ao Paulo},
  addressline={Av.~Trab.~S\~ao Carlense, 400},
  city={S\~ao Carlos},
  postcode={13566-590},
  country={Brazil}
}   

\begin{abstract}
We present a strong theoretical foundation that frames a well-defined family of outer-totalistic network automaton models as a topological generalisation of binary outer-totalistic cellular automata, of which the Game of Life is one notable particular case. These ``Life-like network automata'' are quantitatively described by expressing their genotype (the mean field curve and Derrida curve) and phenotype (the evolution of the state and defect averages). After demonstrating that the genotype and phenotype are correlated, we illustrate the utility of these essential metrics by tackling the firing squad synchronisation problem in a bottom-up fashion, with results that exceed a $90\%$ success rate.
\end{abstract}



\begin{keyword}
Network automata \sep Random Boolean networks \sep Cellular automata \sep The Game of Life \sep Graph theory \sep Firing squad synchronisation problem


\end{keyword}

\end{frontmatter}



\section{Introduction}
\label{sec:introduction}

\noindent A mosaic of intricate patterns can arise from the iterated execution of very simple actions, and nothing demonstrates this emergence of complexity more elegantly than cellular automata (CAs) \cite{wolfram1984universality}. The study of CAs has a long history \cite{sarkar2000brief} throughout which many different varieties and applications have been studied \cite{bhattacharjee2020survey,rollier2025comprehensive}, but the basic setup of this discrete dynamical model is simple to describe. Consider a grid of cells in which every cell can take one of two states. In subsequent time steps, the state of each cell is updated following a `local update rule'. This rule simply maps the states of cells in a well-defined `neighbourhood' of each cell at time step $t$ to the state of that cell in time step $t+1$. This CA is discrete in time, space and state, all cells follow the same deterministic rule, and all cells are updated synchronously. The most well-known example of such a CA is the Game of Life \cite{gardner1970mathematical} and its many variants \cite{adamatzky2010game,evans2003larger}, which follow a so-called `outer-totalistic' local update rule, where a cell's state update only depends on its current state and the state sum of its neighbouring cells.\\

Here we study a family of binary outer-totalistic CAs whose cells (or `nodes') live on a general network, rather than a regular grid. As we indicate in \cite{rollier2025comprehensive}, this topological generalisation effectively brings us to the study of Boolean networks \cite{aldana2003boolean,drossel2008random,schwab2020concepts} with a uniform local update rule, which is a particular type of network automaton (NA). Several researchers have contributed to the exploration of NAs \cite{watts1999global,tomassini2005evolution,marr2009outer,gog2012dynamics,macedo2013dynamic,goles2021generating}, and for uniform binary outer-totalistic NAs in particular, Marr and H\"utt have made important advances regarding the relationship between topology and dynamics \cite{marr2009outer,marr2005topology,marr2012cellular}. In turn, these results have been mobilised in applications including network identification \cite{miranda2016exploring,zielinski2024network}, image classification \cite{ribas2020life}, and authorship attribution tasks \cite{machicao2018authorship}. In particular, these applications take Life-like CAs (LLCA) \cite{adamatzky2010game,pena2021life} and impose a topological generalisation, creating so-called Life-like NAs (LLNAs).\\

We anticipate that we are far from tapping the full potential of these intriguing systems, in the sense that the rich diversity found within the dynamical behaviour of LLNAs warrants more applications in a wider field, while their elegant simplicity should allow for a deeper understanding than is currently achieved. The reason for this is primarily that LLNAs have not yet been sufficiently defined and theoretically studied, such that we lack the essential tools to predict and describe the model's emergent behaviour. In this contribution, we fill that gap by providing the metrics to quantify and connect -- using Kauffman's phrase \cite{kauffman1990requirements} -- the `genotype' and `phenotype' of LLNAs.\\

We first define a generalised notion of LLNAs which obeys some desired symmetries (Sec.~\ref{sec:definition}). Second, we investigate the model by considering properties of the local update rule itself, i.e.~genotype parameters (Sec.~\ref{sec:genotype-parameters}). This involves understanding the Hamming weight and mean-field theory, and understanding the Boolean sensitivity and the Derrida plot. Third, we investigate the model by considering properties of the evolved system, i.e.~phenotype parameters (Sec.~\ref{sec:phenotype-parameters}). In particular we show how the average state over the system evolves, and how an initial defect spreads. We calculate and compare these parameters for various topologies, and demonstrate how the genotype reveals information about the phenotype. Fourth, we give a practical example of how we can use these metrics to find an LLNA rule that exhibits some desired properties (Sec.~\ref{sec:fssp}), particularly a rule that excels at solving the firing squad synchronisation problem \cite{moore1968generalized}.


\section{Definition}
\label{sec:definition}


\noindent Outer-totalistic NAs generalise the grid topology of CAs to any kind of network \cite{marr2009outer}. Because generally not every node has the same degree, a node's state update may depend on a varying number of neighbouring nodes. A uniform local outer-totalistic update rule must therefore be of the general shape $\phi = \phi(s_i, \rho_i)$. Here $s_i = s(v_i)$ is the state of node $v_i$ (also known as a vertex), and $\rho_i = \frac{1}{k_i}\sum_{j}A_{ij}s_j$ is the density (state average) over all $k_i$ nodes in $v_i$'s neighbourhood, indicated by the sum over the adjacency matrix elements $A_{ij}$. This is a function $\phi : \{0,1\}\!\times\![0, 1] \to \{0,1\} : (s_i, \rho_i) \mapsto \phi(s_i, \rho_i)$, or more explicitly:
\begin{equation}
\label{eq:local-update-rule}
    \phi(s_i, \rho_i) = \begin{cases}
        1, \qquad \text{if } s_i = 0 \text{ and } \rho_i \in \bigcup B,\\
        1, \qquad \text{if } s_i = 1 \text{ and } \rho_i \in \bigcup S,\\
        0, \qquad \text{else}.
    \end{cases}
\end{equation}
Here $B$ and $S$ are sets of subintervals of the unit interval, which we will respectively call the `born' set and the `survive' set, using terminology from the Game of Life \cite{adamatzky2010game}. We can further characterise these sets by considering that the function $\phi$ should reduce to the outer-totalistic CA update rule when the topology is a regular lattice. We therefore choose $B, S \subseteq R$, where $R$ is a set of $r$ equally-spaced pairwise disjoint intervals whose union is $[0,1]$. For LLCAs, the `resolution' parameter must be $r=9$, allowing for a bijective map between the intervals in $R$ and the possible sums of living cells directly or diagonally adjacent to the central cell (i.e.~the Moore neighbourhood). However, in the context of LLNAs there is no topological reason to prefer one $r$ value over another.\\

One symmetry consideration imposes a further restriction on the boundaries of the intervals in $R$, as the LLNA ought to respect the same symmetry operations as CAs do \cite{wolfram1983statistical}. Left-right symmetry and rotational symmetry are undefined in networks, but are respected by default in outer-totalistic systems. Symmetry by complementation (switching all $0$s and $1$s) is not, however, because $\rho_i$ can take any rational value in the unit interval. This imposes the subtle requirement that all subsets in $R$ are mapped to \textit{other subsets} upon complementation, i.e.~mirrored around $\rho_i = \frac{1}{2}$. Concretely, the elements of $R$ should therefore be of the form
\begin{align}
    R   &= \{R_0, R_1, \ldots, R_{r-1}\}\notag\\
        &= \left\{\left[0, \frac{1}{r}\right[, \left[\frac{1}{r}, \frac{2}{r}\right[, \ldots, \left[\frac{(r-1)}{2r}, \frac{(r+1)}{2r}\right], \ldots,\right.\label{eq:r-intervals}\\
        & \left.\qquad\qquad \left]\frac{r-2}{r}, \frac{r-1}{r}\right], \left]\frac{r-1}{r}, 1\right]\right\}. \notag
\end{align}
Note how every element is a half-open interval except for the middle one, which requires that the resolution $r$ be an odd integer. With this definition, Eq.~\eqref{eq:local-update-rule} respects the principles of homogeneity, isotropy, and functional simplicity outlined by Marr and H\"utt \cite{marr2009outer}, and adds the fundamental principle of complementation symmetry \cite{wolfram1983statistical}.\\



\begin{figure}
    \centering
    \includegraphics[width=\linewidth]{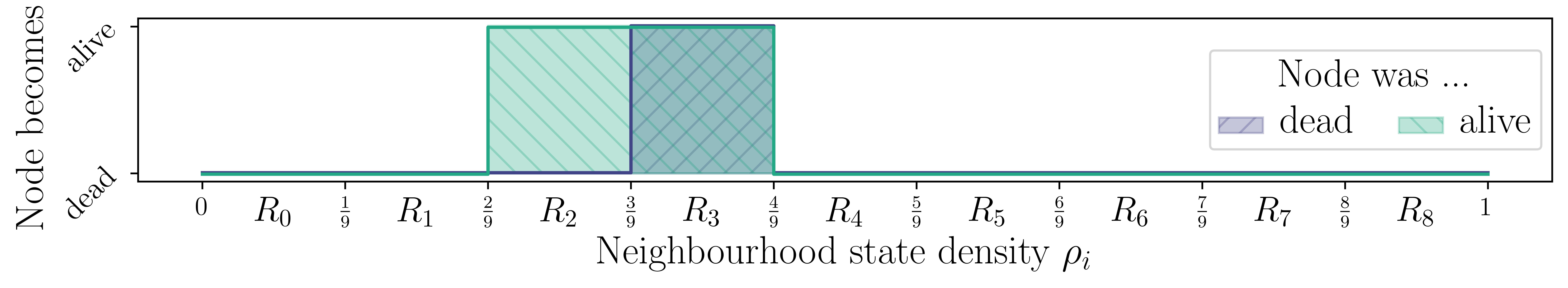}
    \caption{This diagram represents the local update rule of the Game of Life, phrased in the formalism of an LLNA. The horizontal axis represents the state density of the neighbourhood. The vertical axis represents the response of the local update function, i.e.~whether the central cell will be alive ($1$) or dead ($0$) in the next time step. The horizontal axis is divided into nine density intervals, representing the set $R$ whose elements are defined in Eq.~\eqref{eq:r-intervals}. The colours indicate whether the central cell is dead or alive in the current time step, which determines whether it should respond to the $B$ set or the $S$ set in Eq.~\eqref{eq:local-update-rule}. Following the notation of Eq.~\eqref{eq:compact-notation}, this rule is denoted as $\phi^9_{8,12}$.}
    \label{fig:gol_as_llna-diagram}
\end{figure}

In Eq.~\eqref{eq:r-intervals}, each odd natural number $r$ generates a set of rules $\Phi^r$ with cardinality $2^{2r}$. Within $\Phi^r$, a particular local update rule is defined by the choices $B, S \subseteq R$. This definition encompasses the classical threshold rules on networks \cite{marr2005topology}, and -- if the network is a simple grid -- all possible LLCAs \cite{adamatzky2010game}. For the Game of Life, for example, we find $r = 9$, $B = \{R_3\}$, $S = \{R_2, R_3\}$. We will compactly write the local update rule as $\phi^r_{\beta,\sigma}$ or simply $\phi$ if the context is clear or particular values of the $r, \beta, \sigma$ values are irrelevant. Here, the subscripts $\beta$ and $\sigma$ are decimals, inspired by the notation (``code'') for totalistic CAs \cite{wolfram1984universality}:
\begin{equation}
    \beta = \sum_{i : R_i \in B} 2^i, \qquad \sigma = \sum_{i : R_i \in S} 2^i,
    \label{eq:compact-notation}
\end{equation}
such that $0 \leq \beta, \sigma < 2^r$. For example, we write the local update rule of the Game of Life as $\phi^9_{8,12}$ (Fig.~\ref{fig:gol_as_llna-diagram}), and Wolfram rule $150$ as $\phi^3_{2,5}$.\\

By design, every rule has an equivalent rule that generates identical behaviour in the NA when complementing all binary states in the initial configuration. It can easily be shown that we have an equivalence between local update rules $\phi^r_{\beta,\sigma}$ and $\phi^r_{\beta',\sigma'}$ upon state inversion, where simultaneously
\begin{equation}
    \label{eq:equivalent_rules}
    \beta \sim \beta' = 2^r - 1 - \sum_{i : R_i \in S} 2^{r-1-i},
    \qquad
    \sigma \sim \sigma' = 2^r - 1 - \sum_{i : R_i \in B} 2^{r-1-i}.
\end{equation}
For example, rule $\phi^5_{11,19}$ is equivalent to $\phi^5_{5,6}$, which means that they display exactly the same dynamical behaviour, but simply with inverted colours. A rule is equivalent to itself if $\beta = \beta'$ and $\sigma = \sigma'$ (e.g.~rule $\phi^5_{6,19}$), which is the case for $2^r$ combinations of $\beta$ and $\sigma$. Therefore, we have $2^{2r-1} + 2^{r-1}$ non-equivalent rules for every resolution $r$, i.e.~a little over half of all rules.



\section{Genotype parameters}
\label{sec:genotype-parameters}

\noindent As the number of available rules grows exponentially with the resolution $r$, it is useful to characterise and classify these rule by means of quantitative parameters. The reader may find an excellent overview of genotype and phenotype parameters for CAs in \cite{vispoel2022progress}; we will generalise two essential parameters for NAs. Ideally, a genotype parameter exhibits the eight desirable properties listed by de Oliveira et al.~\cite{oliveira2000guidelines}. Arguably the most important of these properties is a correlation between the parameter of the local update rule and the resulting dynamical behaviour, which we will turn our attention to in Subsec.~\ref{subsec:relationship-state-vs-defect-average}.\\

\begin{figure}
    \centering
    \includegraphics[width=\linewidth]{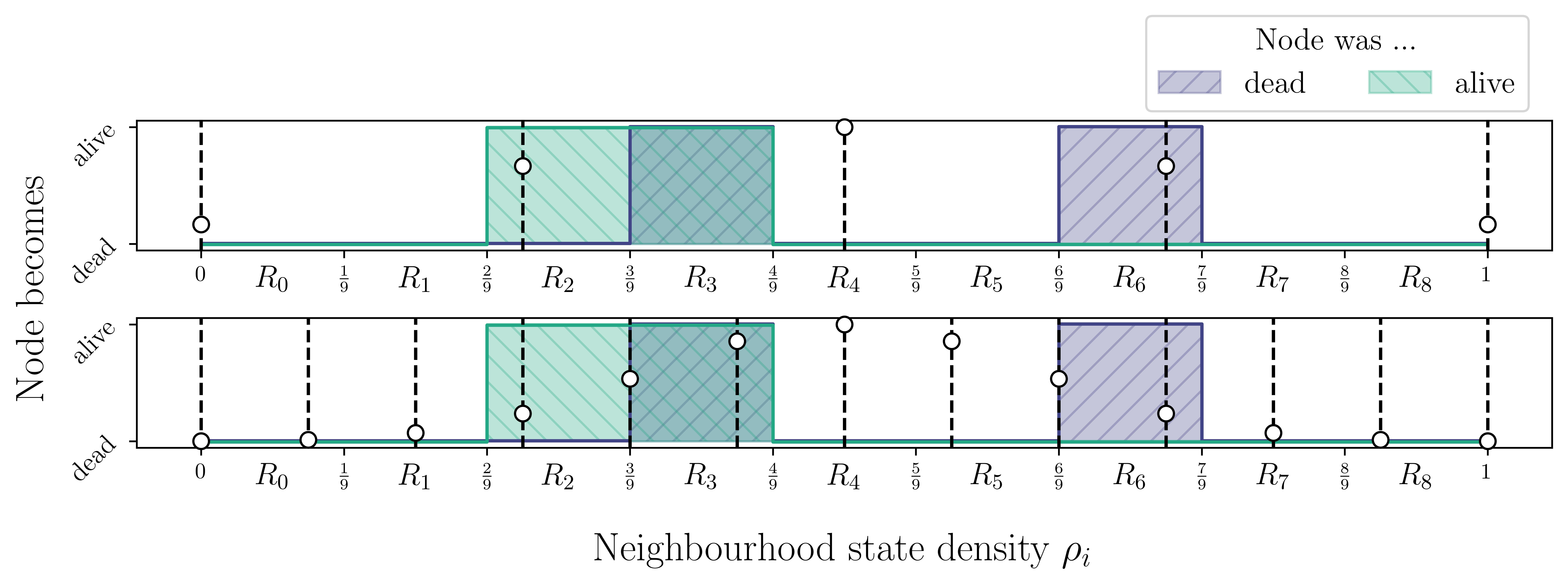}
    \caption{Local update rule diagram for $\phi^9_{72,12}$ (known as the LLCA ``HighLife'' \cite{eppstein2010growth}) for node degrees $4$ (top) and $12$ (bottom). The vertical dashed lines indicate the possible state densities for the respective degree, and the white circles indicate the binomially distributed relative abundance of the configurations that result in this density. \textit{Top}: node $v_i$ has degree $k_i=4$, such that only five densities can be taken, altogether skipping intervals $R_1, R_3$, $R_5$ and $R_7$. \textit{Bottom}: for degree $k_i=12$, every density interval can be reached, but most of the configurations result in a density in $R_3$, $R_4$ or $R_5$. This results in an average behaviour that is significantly different from the degree-$4$ case (cf.~Fig.~\ref{fig:mean_field_curve-derrida_curve-HW_BS_distribution-R9B72S12}).}
    \label{fig:compare-two-degrees_diagram}
\end{figure}

Here we will first discuss the Hamming weight and the Boolean sensitivity \cite{shmulevich2004activities}. We demonstrate their intimate relation to mean-field theory \cite{schulman1978statistical} and the Derrida plot \cite{derrida1986evolution} for LLNAs. While this relationship is valid for any rule in $\Phi^r$, we will use rule $\phi^9_{72,12}$ as an example to visually illustrate these quantities. This rule is the topological generalisation of the well-studied Life-like CA called ``HighLife'' \cite{eppstein2010growth}. Fig.~\ref{fig:compare-two-degrees_diagram} depicts its update diagram in, overlaid with the distribution of possible neighbourhood state densities $\rho_i$ for two different degrees of $v_i$.

\subsection{Hamming weight and mean-field theory}

\subsubsection{Hamming weight}


\noindent The (normalised) Hamming weight (HW) is simply the fraction of local update rule inputs that map to a living node. Note that this is mathematically and conceptually identical to the Langton parameter for elementary CAs, which is a reasonably good predictor for dynamical behaviour \cite{langton1990computation}. For our model, the HW for a given update function $\phi$ and a network with degree distribution $P(k)$ is
\begin{align}
    \label{eq:hamming_weight}
    \text{HW} = \sum_{k}P(k)\text{HW}_k = \sum_{k}P(k)\frac{1}{2^{k+1}}\!\!\sum_{s \in \{0,1\}}\sum_{q=0}^{k} \binom{k}{q}\,\phi\!\left(s, \frac{q}{k}\right).
\end{align}
The outer sum is over all degrees $k$ in the network, and is weighted by the degree distribution. The middle sum is over the states of a particular node, and the inner sum over the possible values $q$ representing the sum of the states of its neighbouring nodes. The binomial factor $\binom{k}{q}$ accounts for the total number of neighbourhood configurations that result in the same state density $q/k$, and the division by $2^{k+1}$ accounts for all possible configurations of a node and its neighbourhood.

\subsubsection{Mean-field curve}

\begin{figure}
    \centering
    \includegraphics[width=1\linewidth]{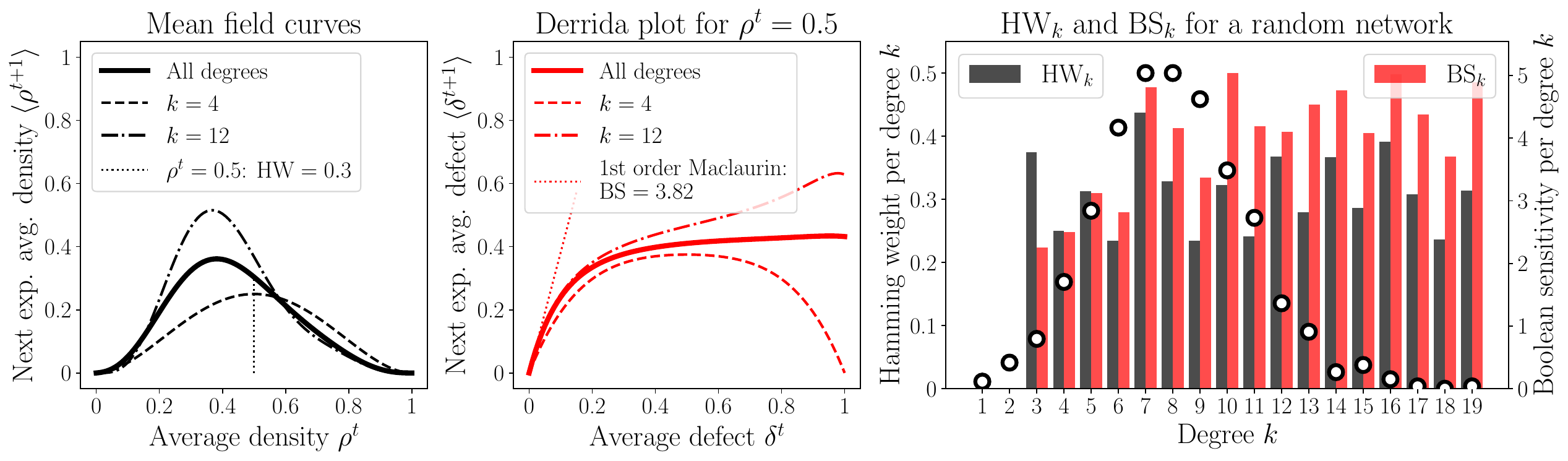}
    \caption{Genotype parameters for rule $\phi^9_{72,12}$ on a random network with $N=900$ nodes and $3600$ edges. \textit{Left}: mean-field curve for degrees $4$ (dashed black) and $12$ (dash-dotted black), and weighted (solid black) by the degree distribution. The associated HWs per degree are mapped from $\rho^0=0.5$ (dotted black): $\text{HW}_4 = 0.25$, $\text{HW}_{12}=0.37$, and $\text{HW}=0.30$. \textit{Middle}: Derrida plot for average density $\rho^0=0.5$, for degrees $4$ (dashed red) and $12$ (dash-dotted red), and weighted (solid red) by the degree distribution. The associated Boolean sensitivities per degree correspond to the slope of the tangent at the origin (dotted red): $\text{BS}_4 = 2.50$, $\text{BS}_{12}=4.10$, and $\text{BS}=3.82$. \textit{Right}: all degree-specific HWs (black bars) and Boolean sensitivities (red bars). The relative abundance of the degrees in the network (white circles) indicates how degree-specific values are weighted.}
    \label{fig:mean_field_curve-derrida_curve-HW_BS_distribution-R9B72S12}
\end{figure}

\noindent In mean-field theory \cite{schulman1978statistical}, the mean-field curve expresses the expectation value of the average of all states in the next time step, neglecting local topological structure \cite{gutowitz1987local}. For a local update function $\phi$ and some random configuration with global state average $\rho^t$ (now without index $i$), we find that the expected average over all states at time step $t+1$ is
\begin{align}
    \label{eq:mean_field_curve}
    \langle\rho^{t+1}\rangle = \frac{1}{N}\sum_{i=1}^N \langle s^{t+1}_i \rangle = \sum_{k}P(k)\!\!\sum_{s\in\{0, 1\}}\!\!P(s | \rho^t)\sum_{q=0}^{k} P(q | k, \rho^t)~\phi(s, q/k).
\end{align}
Let us break this down. The expected average $\langle\rho^{t+1}\rangle$ is identical to the average over all expectation values of the state $\langle s_i^{t+1} \rangle $ of all nodes in the network $\{v_i~|~1\leq i \leq N\}$. This average of expectation values depends on a triple sum. The outermost sum is again over the degree distribution. The middle sum is over the states $s$ of a particular node, and is weighted by the probability $P(s|\rho^t)$ of finding this state, provided that the configuration at time $t$ has a state average of $\rho^t$. The inner sum is over all possible living nodes $q$ in that node's neighbourhood, and is weighted by the probability $P(q | k, \rho^t)$, again given $\rho^t$, and provided that the neighbourhood has size $k$. The aforementioned probabilities follow simple binomial distributions if we ignore local topological structure:
\begin{align*}
    P(s | \rho^t)       &= (\rho^t)^s (1-\rho^t)^{1-s},\\
    P(q | k, \rho^t)    &= \binom{k}{q} (\rho^t)^q (1-\rho^t)^{k-q}.
\end{align*}
If $\rho^t = 1/2$, all neighbourhood configurations are equally likely, and we find that the expected state average $\langle\rho^{t+1}\rangle$ is equal to the HW, as Eq.\eqref{eq:mean_field_curve} reduces to Eq.~\eqref{eq:hamming_weight}. We show an example for rule $\phi^9_{72,12}$ operating on a random network Fig.~\ref{fig:mean_field_curve-derrida_curve-HW_BS_distribution-R9B72S12}, displaying the effect of the non-uniform node degree on the system's mean-field response (left) and on the degree-specific HWs (right).

\subsection{Boolean sensitivity and the Derrida plot}

\subsubsection{Boolean sensitivity}

\noindent The Boolean sensitivity (BS) quantifies the update rule's sensitivity to toggling an input state ($0 \leftrightarrow 1$). More precisely, it is the number of Hamming neighbours of a particular input configuration resulting in a different output than the original configuration, weighted over all possible configurations. The BS is therefore equivalent to what has been dubbed $\mu$-sensitivity \cite{binder1993phase} for elementary CAs, and is strongly associated with the concept of Boolean derivatives \cite{vichniac1990boolean}. Additionally, it checks all eight boxes for being an appropriate rule table parameter \cite{oliveira2000guidelines}. For degree $k$ we find
\begin{align}
    \text{BS}_k = \frac{1}{2^{k+1}}\!\!\sum_{s \in \{0,1\}} \sum_{q=0}^{k}\binom{k}{q} \Big( \text{IS}(s, q, k) + \text{NS}(s, q, k)\Big),
    \label{eq:boolean-sensitivity}
\end{align}
where we may again average over the network: $\text{BS} = \sum_k P(k)\text{BS}_k$. We make a distinction between the `identity sensitivity' IS, and the `neighbourhood sensitivity' NS, which we shall see intuitively reflects the outer-totalistic nature of our local update rule. The IS quantifies whether the output $\phi(s_i, \rho_i)$ of the local update rule changes when the state $s_i$ of the node $v_i$ is toggled, hence measuring the sensitivity to its own identity of the map to the next time step. If we take $\oplus$ to be the \textsc{xor} operator, we then find
\begin{align*}
    \text{IS}(s, q, k) = \phi\left(s, \frac{q}{k}\right) \oplus \phi\left(s \oplus 1, \frac{q}{k}\right).
\end{align*}
On the other hand, the NS quantifies whether the state of the node $v_i$ in the next time step is sensitive to a minimal increase or decrease of the density of states of its neighbouring nodes:
\begin{align*}
    \text{NS}(s, q, k) = q\text{NS}^-(s, q, k) + (k-q)\text{NS}^+(s, q, k),
\end{align*}
where we differentiate between a decrease ($\text{NS}^-$) or increase ($\text{NS}^+$) of the density as a result of a bit flip, and multiply by the number of choices for this bit flip ($q$ resp.~$k-q$). In other words, we define
\begin{align*}
\text{NS}^\pm(s, q, k) &= \phi\bigg(s,\frac{q} {k}\bigg) \oplus \phi\bigg(s,\frac{q\pm1}{k}\bigg).
\end{align*}

This distinction between IS and NS enables a clear visual interpretation. Again referring to Fig.~\ref{fig:compare-two-degrees_diagram}, the IS is the number of configurations that fall within a density interval that has a single colour (not both and not none). The NS corresponds to the `jaggedness' of the born and survive sets of density intervals and essentially counts how many times the diagram changes its response when hopping between adjacent state densities. If $B$ (or $S$) consists mainly of contiguous intervals, the visual `jaggedness' of the diagram will be low, as will be the NS value.

\subsubsection{Derrida plot}

\noindent The Derrida plot \cite{derrida1986evolution} is related to the BS much like the mean-field curve is related to the HW. It describes how a defect, defined as the bitwise \textsc{xor} operation on two nearly identical configurations, initially propagates through the network \cite{fretter2009perturbation}. More specifically, it maps the normalised Hamming distance $\delta^t$ between two configurations at time step $t$ to the expected normalised Hamming distance $\langle\delta^{t+1}\rangle$ between these configurations after evolving both one single time step. This map can be formulated analytically (albeit quite tediously):
\begin{align}
    \langle\delta^{t+1}\rangle &= \frac{1}{N} \sum_{i=1}^N \langle d_i^{t+1} \rangle = \sum_k P(k)\!\! \sum_{s \in \{0,1\}}\!\!\! P(s|\rho^t)
                \sum_{q=0}^k P(q|k,\rho^t)\!\! \sum_{c \in \{0,1\}}P(c|\delta^t) \ldots \notag \\
             &\quad \ldots \times
                \sum_{d=0}^k P(d|k, \delta^t)
                \sum_{\tau} P(\tau|k,d,q) \, \mathcal{D}(k, s, q, c, d, \tau).
    \label{eq:derrida-map}
\end{align}
This expression indicates that the expected normalised Hamming distance at time $t+1$ is identical to the average over all expectation values of the defect $\langle d_i^{t+1}\rangle$ of node $v_i$, for all nodes in the network. This is identical to a sixfold sum. The outer three sums are familiar from Eq.~\eqref{eq:mean_field_curve}. The fourth sum counts the probabilities that the cental node $v_i$ is toggled ($c=1$) or not ($c=0$), given that the normalised Hamming distance between both configurations at time step $t$ is $\delta^t$. The fifth sum adds the probabilities that $d$ neighbourhood nodes are toggled at all (whatever their state value), given $\delta^t$ and neighbourhood size $k$. The innermost sum adds the probabilities that $\tau$ of the $d$ toggles are `killer toggles' (bit flips that turn a $1$ into a $0$), given $k$, $d$, and the number of living neighbourhood nodes $q$, so
\begin{align*}
    \mathcal{D}(k, s, q, c, d, \tau) = \phi\bigg(s, \frac{q}{k}\bigg) \oplus \phi\bigg(s\oplus c, \frac{q-2\tau+d}{k}\bigg).
\end{align*}
The formula above expresses the defect compared to the default state at the \textit{next} time step, after the relevant changes to a node and its neighbourhood sum are made at the \textit{current} time step. If we again ignore local topological structure, the probabilities follow respectively two binomial distributions and one hypergeometric one:
\begin{align*}
    P(c|\delta^t)   &= (\delta^t)^c (1-\delta^t)^{1-c},\\
    P(d|k,\delta^t) &= \binom{k}{d}(\delta^t)^d (1-\delta^t)^{k-d}, \\
    P(\tau|k, d, q)    &= \binom{q}{\tau}\binom{k-q}{d-\tau} \bigg/ \binom{k}{d}.
\end{align*}
In the latter distribution $\max(0, d+q-k) \leq \tau \leq \min(d,q)$.\\

As can be inferred from Eq.~\eqref{eq:derrida-map}, there is a distinct Derrida plot for every value of the average state density $\rho^t$. For $\rho^t=0.5$, we invite the reader to verify that the BS is identical to the first-order Taylor coefficient of the Maclaurin expansion in $\delta^t$. This conveys that the BS value is equal to the slope of the tangent through the origin of the Derrida plot, which quantifies the system's sensitivity to small perturbations. Note that the logarithm of this slope is known as the Derrida coefficient, which is positively correlated to a system's chaoticity \cite{harris2002model}. In Fig.~\ref{fig:mean_field_curve-derrida_curve-HW_BS_distribution-R9B72S12}, we show an example of a Derrida plot (middle) and some $\text{BS}_k$ values (right), again for rule $\phi^9_{72,12}$ on a random network.

\subsection{Relationships, limits and symmetries of HW and BS}
\label{subsec:relationships-limits-symmetries}

\noindent We close this section with three final remarks. First as indicated in Ref.~\cite{shmulevich2004activities}, the HW and BS values must be related. If, for example, $\text{HW}_k$ is low for a particular degree $k$, the function maps to $0$s for most inputs. In that case, the probability of encountering a different output upon flipping a neighbourhood state is low, i.e.~the $\text{BS}_k$ is low as well.\\

Second, while calculating degree-specific HW or BS values is typically not straightforward, the limit for high degree is easy to estimate. The binomial distribution of the number of configurations that result in a particular state average $\rho_i$ will approach a normal distribution when $k \to \infty$, so the HW will be dominated by the (non-)activation of $R_i$ intervals close to $\rho_i = 0.5$. The BS, on the other hand, will always tend towards the identity sensitivity related to these central intervals, as the neighbourhood sensitivity vanishes for large degrees. These limit considerations allows one to make accurate characterisations of highly connected networks.\\


Third, we highlight that our definition of outer-totalistic network automata results in additional symmetries for HW and BS values. We list the operations under which the BS and HW values are (quasi) invariant in Tab.~\ref{tab:operations-identical-HW-BS}. Moreover, by combining operations in this table, we see that equivalent rules -- for which $(B,S) \to (\bar{S}^\text{C}, \bar{B}^\text{C})$, cf.~Eq.~\eqref{eq:equivalent_rules} -- have an identical BS, and a complementary HW. This is of course expected from complementation symmetry.


\begin{table}
    \centering
    \caption{Operations that do not affect the BS value. Above the dashed line, also the HW value is invariant. Under the dashed line, the HW value is mapped to its complement $\text{HW} \mapsto 1-\text{HW}$, as toggling the outputs generates dual rules. $B^\text{C}$ is short for $R \setminus B$, and $\bar{B}$ is identical to the $B$ set after state complementation (for which all densities are mapped $\rho \mapsto 1-\rho$).}
    \begin{tabular}{lL{5cm}L{3.8cm}}
        Map & Input/output manipulation & Diagram \\ \hline
        $B \longleftrightarrow S$ & Toggle node state $s_i$ & Switch colours \\
        $(B, S) \mapsto (\bar{B}, \bar{S})$ & Toggle nodes neighbouring to $v_i$ & Mirror intervals around $\rho=1/2$ \\ \hdashline
        $(B, S) \mapsto (B^\text{C}, S^\text{C})$ & Toggle outputs & Toggle all intervals \\
    \end{tabular}
    \label{tab:operations-identical-HW-BS}
\end{table}


\section{Phenotype parameters}
\label{sec:phenotype-parameters}

\noindent The complement of genotype parameters like the HW and BS are parameters that quantify the \textit{actual} behaviour of the system, i.e.~phenotype parameters. The phenotype comprises \textit{a posteriori} metrics that quantify the effect of the update mechanism by studying subsequent state configurations. The calculation of the phenotype therefore requires the evolution of this configuration.\\

Most relevant phenotype parameters may be found in Vispoel et al.'s survey \cite{vispoel2022progress}, and popular parameters for NA research include the Shannon and word entropy \cite{marr2009outer, marr2012cellular}, the Lempel-Ziv complexity, and the Lyapunov exponent \cite{miranda2019spatially,vispoel2024damage}. In what follows, we will inspect the phenotype by looking at two simple time series which are the natural counterparts of the genotype parameters discussed earlier. These are the global state average $\rho^t$ and the defect average $\delta^t$, defined respectively as
\begin{align}
    \rho^t = \frac{1}{N}\sum_{i=1}^N s(v_i, t), \qquad \delta^t = \frac{1}{N}\sum_{i=1}^N \Big( s(v_i, t) \oplus s'(v_i, t) \Big).
\end{align}
Here $s(v_i,t)$ is the state of node $v_i$ at time step $t$, and $s'(v_i, t)$ is the state of that node at $t$, where the initial configuration is perturbed in exactly one node. Below we shall demonstrate the possibly strong influence of the initial configuration and the topology on the system's emergent behaviour, i.e.~cases where local topological structure \textit{cannot} be ignored.

\subsection{Computational pipeline}
\label{subsec:computational-pipeline}

Here we briefly consider the experimental setup that we will refer to throughout the remainder of this article. We shall compare time series $\rho^t$ and $\delta^t$ for automata evolved on three simple undirected networks, starting from different initial state averages $\rho^0 \in \{\frac{1}{4}, \frac{2}{4}, \frac{3}{4}\}$. The networks represent an increasing departure from the default topology for Life-like CAs. The first is a toroidally wrapped regular lattice with node degree $8$, i.e.~Life's natural habitat. The second is a small-world network that is generated by applying the Watts-Strogatz algorithm \cite{watts1998collective} onto the aforementioned regular lattice, with rewiring probability $p=0.2$. The third is simply a random network. All these networks have $N=900$ nodes and $3600$ edges. The phenotype is evaluated by considering the median and the interquartile range (IQR) over $900$ samples taken by randomly sampling $30$ initial configurations (with the desired state average) for $30$ realisations of each network type. The LLNAs are evolved over $T=100$ time steps and, where needed, convergence values are found by taking the IQR over the final $\Delta t = 10$ time steps.

\subsection{State average evolution}

\begin{figure}
    \centering
    \includegraphics[width=\linewidth]{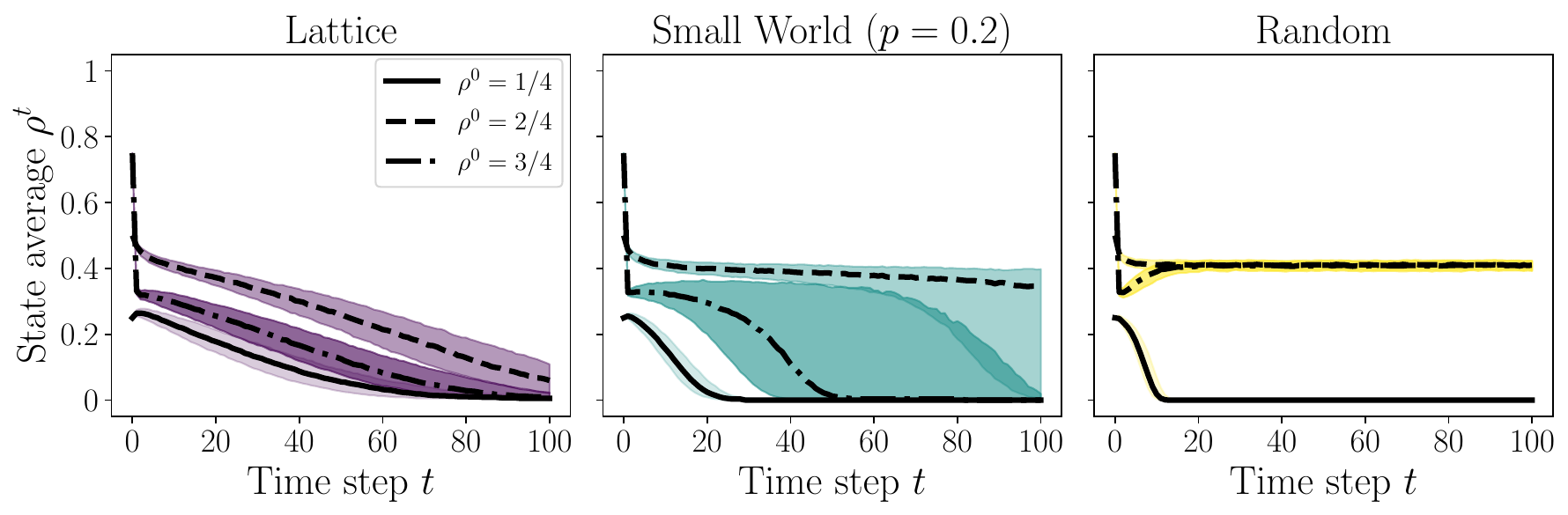}
    \caption{The time series of state averages $\rho^t$ of an NA evolved according to rule $\phi^9_{328,52}$, with initial state density $\rho^0 = 1/4$ (full line), $\rho^0 = 2/4$ (dashed line), or $\rho^0 = 3/4$ (dash-dotted line), for three distinct network topologies. From left to right, these are the toroidal lattice (purple), small-world networks (teal), and random networks (yellow).}
    \label{fig:state-average-evolution_three-networks_R9B328S52}
\end{figure}

\noindent The evolution of state configurations on NAs can be quantitatively \cite{baetens2011topological} or even qualitatively \cite{behrens2024dynamical} dependent on small changes in topology and initial configuration, and this is no different for most rules in the NA family studied here. We show this in Fig.~\ref{fig:state-average-evolution_three-networks_R9B328S52} for rule $\phi^9_{328,52}$ (whose Life-like CA counterpart is called the Morley rule \cite{eppstein2010growth}) because its behaviour clearly illustrates the topological and configurational sensitivity. Note how the $\rho^0$ value will generally affect the time to convergence, where in this case a high value for the initial state average (dash-dotted line) will cause the system to converge faster than an intermediate value (dashed line). Note additionally how the topology affects convergence time: while a toroidal lattice (purple) supports gradual convergence towards a homogeneous state, a small-world network (teal) supports more sudden convergence, and a random network (yellow) sometimes does not enable convergence at all. Such heterogeneous and perhaps counter-intuitive dependencies on initial configuration and topology are commonplace for these systems \cite{tomassini2005evolution,marr2009outer,baetens2013topology}.\\

\begin{figure}
    \centering
    \includegraphics[width=0.8\linewidth]{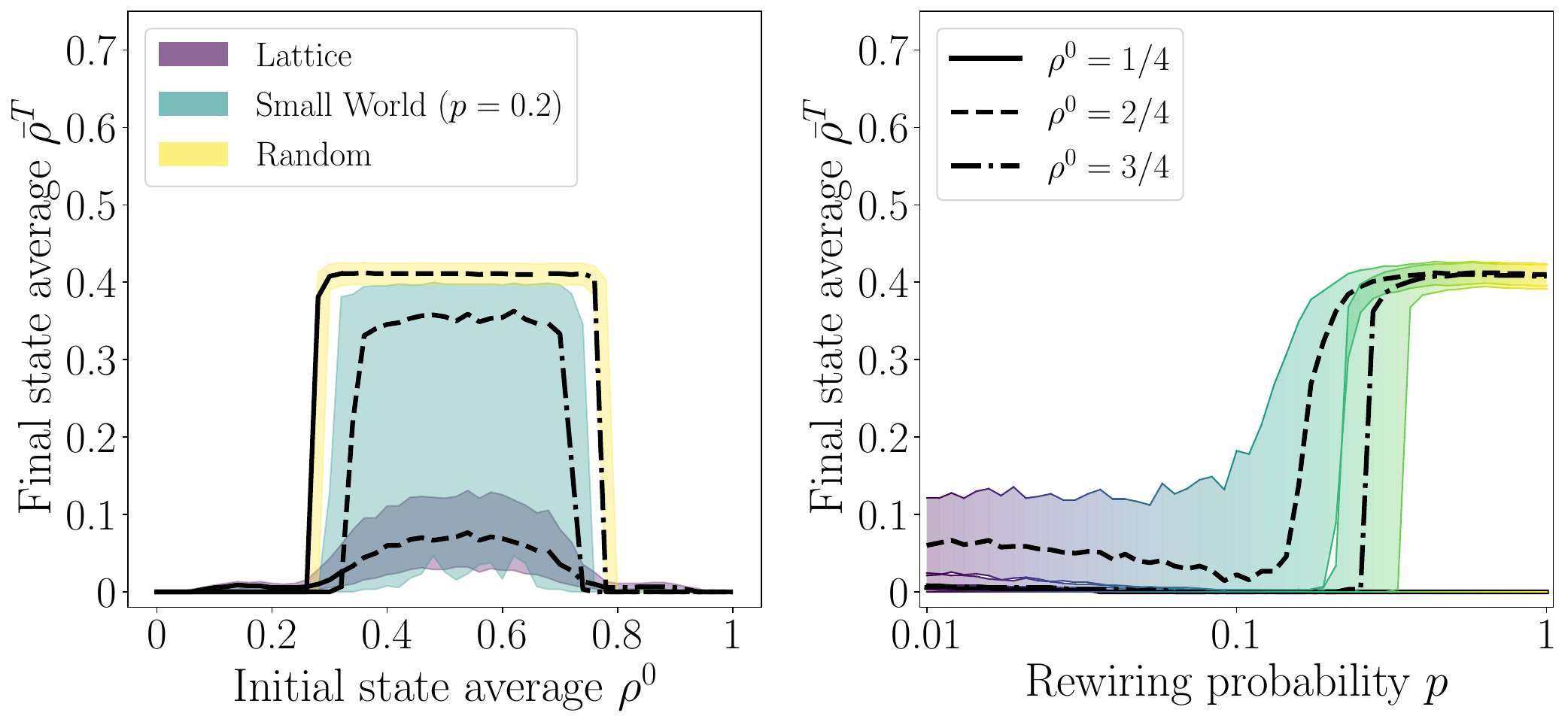}
    \caption{Sweep of dynamical and topological values and their effect on the system's state average convergence, with colours and linestyles corresponding to the time series in Fig.~\ref{fig:state-average-evolution_three-networks_R9B328S52}. \textit{Left}: $3$ networks with $51$ linearly spaced initial state densities. Only intermediate values of the initial state density can result in an LLNA that does not go extinct ($\bar{\rho}^T=0$), with a strong cut-off for random networks. \textit{Right}: $51$ networks were generated using the Watts-Strogatz algorithm with logarithmically spaced rewiring probabilities for three distinct initial state densities. For a high initial density, the system displays a phase transition between $p=0.2$ and $p=0.3$.}
    \label{fig:final_state_densities_vs_init_conf_and_topology_R9B328S52}
\end{figure}

In Fig.~\ref{fig:final_state_densities_vs_init_conf_and_topology_R9B328S52} we consider more systematically how the phenotype of rule $\phi^9_{328,52}$ depends on the LLNA's dynamical and topological properties. We do so by plotting the final state average $\bar{\rho}^T$ for a range of initial state densities $\rho^0$ (left) or rewiring probabilities $p$ of the Watts-Strogatz models (right). In order to better represent the actual phenotype, $\bar{\rho}^T$ is not literally the final state average -- which would simply be $\rho^T$, without the bar --, but rather a median value over all samples in the final $\Delta t$ time steps. We observe a phase transitions, for example (in the right-hand plot) when the initial state density is $3/4$ and the rewiring probability is between $0.2$ and $0.3$, and we find such phase transitions for many rules in $\Phi^r$.

\subsection{Defect average evolution}

\begin{figure}
    \centering
    \includegraphics[width=\linewidth]{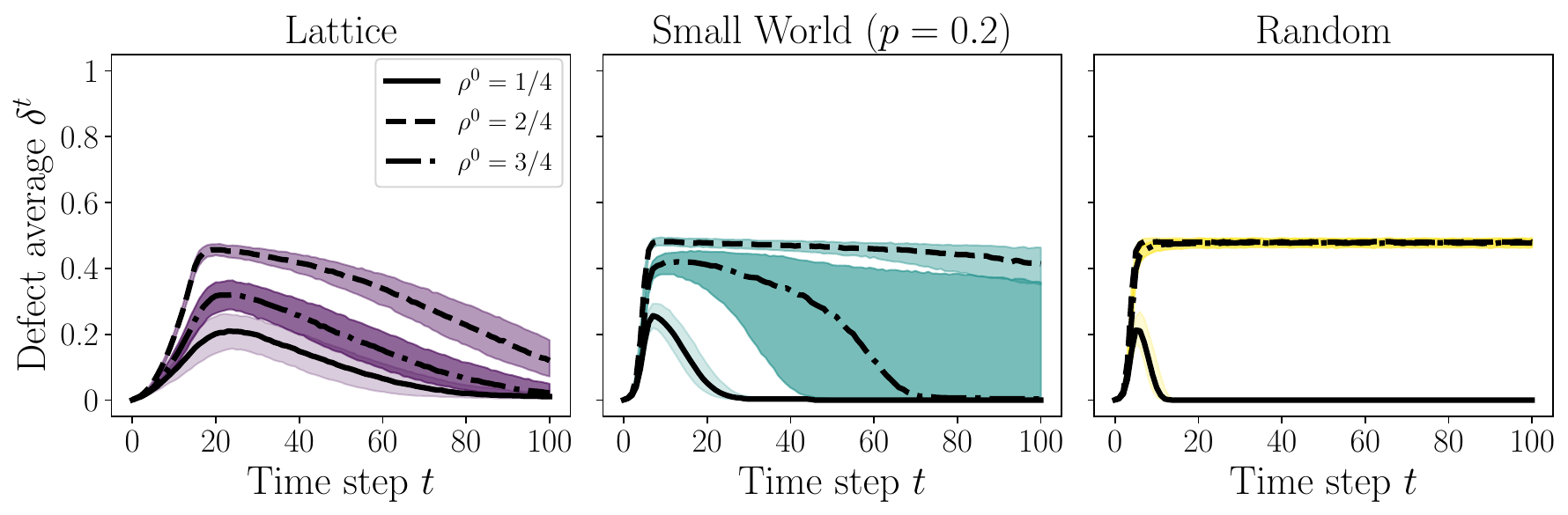}
    \caption{The time series of defect averages $\delta^t$ of an NA evolved according to rule $\phi^9_{328,52}$ for various dynamical and topological scenarios. Compare this to the time series of state averages in Fig.~\ref{fig:state-average-evolution_three-networks_R9B328S52}.}
    \label{fig:defect-average-evolution_three-networks_R9B328S52}
\end{figure}
\begin{figure}
    \centering
    \includegraphics[width=0.8\linewidth]{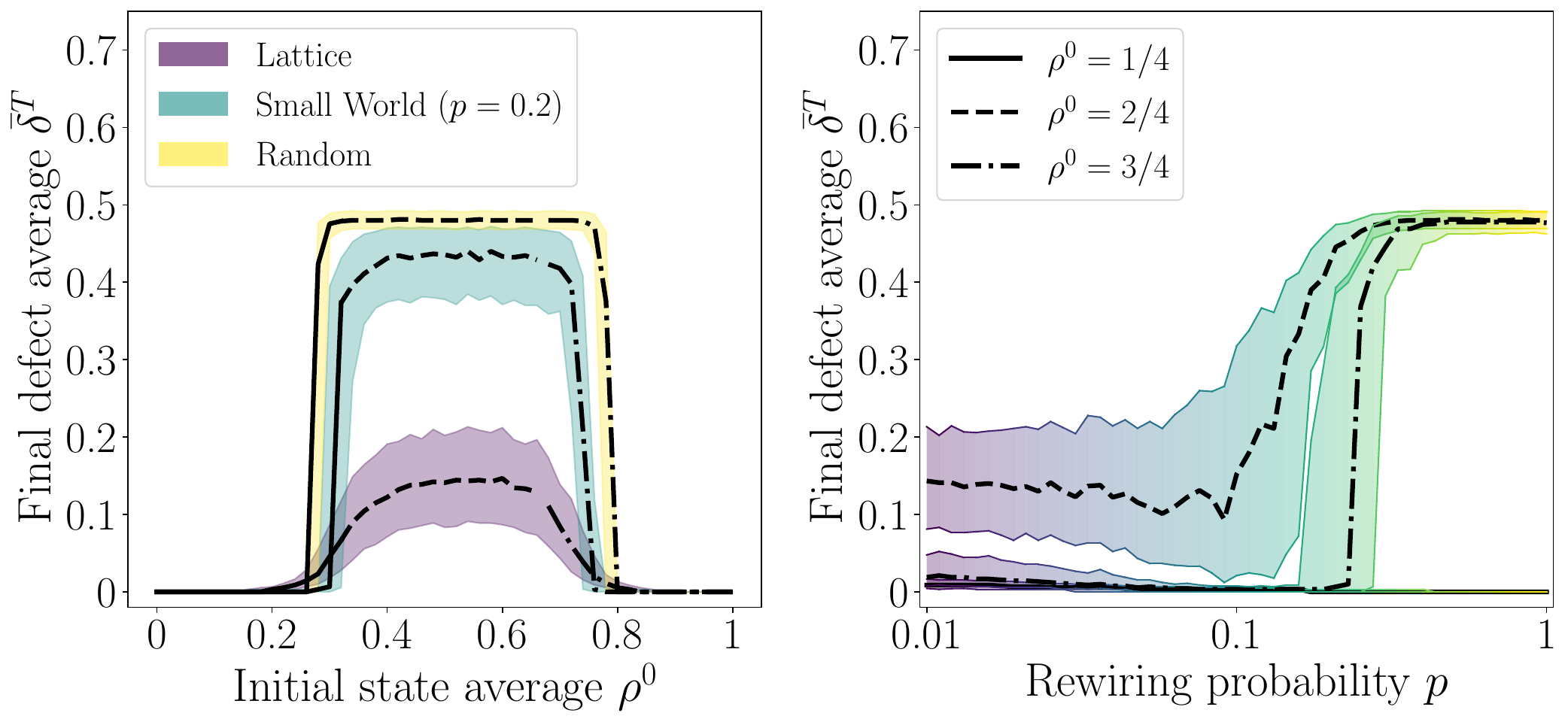}
    \caption{Similar to Fig.~\ref{fig:final_state_densities_vs_init_conf_and_topology_R9B328S52} but now for the final defect density $\bar{\delta}^T$ for rule $\phi^9_{328,52}$. Note that only systems that do not go extinct can propagate defects, so the curves in Fig.~\ref{fig:final_state_densities_vs_init_conf_and_topology_R9B328S52} put constraints on the curves in this figure.}
    \label{fig:final_defect_densities_vs_init_conf_and_topology_R9B328S52}
\end{figure}

\noindent The initial configuration and the topology have been shown to also have a strong influence on how an initially small defect spreads through the network \cite{baetens2013topology,lu2014damage}, and that is also the case for our LLNAs. We again illustrate this by means of rule $\phi^9_{328,52}$: in Fig.~\ref{fig:defect-average-evolution_three-networks_R9B328S52} we show the resulting average defect between both evolved state configurations, which is diverse, indeed. We also take the median quantity $\bar{\delta}^T$ of $\delta^t$ over the final $\Delta t$ time steps, and plot its dependence on the initial state density $\rho^0$ and rewiring probability $p$ in Fig.~\ref{fig:final_defect_densities_vs_init_conf_and_topology_R9B328S52}. The observed discontinuities for particular values of $\rho^0$ and $p$ are characteristic for a phase transition \cite{baetens2013topology,bagnoli2018phase}.

\subsection{Relationship between genotype and phenotype}
\label{subsec:relationship-state-vs-defect-average}

\noindent The genotype parameters are informative about the system right after initialisation, while the phenotype parameters require explicit calculation and express information about the system's entire evolution or final state. Even for random Boolean network models that only slightly diverge from the classic definition, the genotype quickly loses its ability to fully describe the expected dynamics \cite{fretter2009perturbation}, as local topological structure gains importance \cite{gutowitz1987local}. Still, we observe that the genotype and phenotype are intimately related. By comparing Figs.~\ref{fig:state-average-evolution_three-networks_R9B328S52} through \ref{fig:final_defect_densities_vs_init_conf_and_topology_R9B328S52}, it is apparent that there is a connection between the state values and defect values. This can be understood by realising that a defect cannot propagate if the entire network evolves towards a homogeneous state, regardless of the initial configuration. Therefore, if $\bar{\rho}^T$ goes to either $0$ or $1$, the median final state defect $\bar{\delta}^T$ must approach $0$. For an intermediate value of $\bar{\rho}^T$, defect propagation is less easily predicted.\\

\begin{figure}
    \centering
    \includegraphics[width=.85\linewidth]{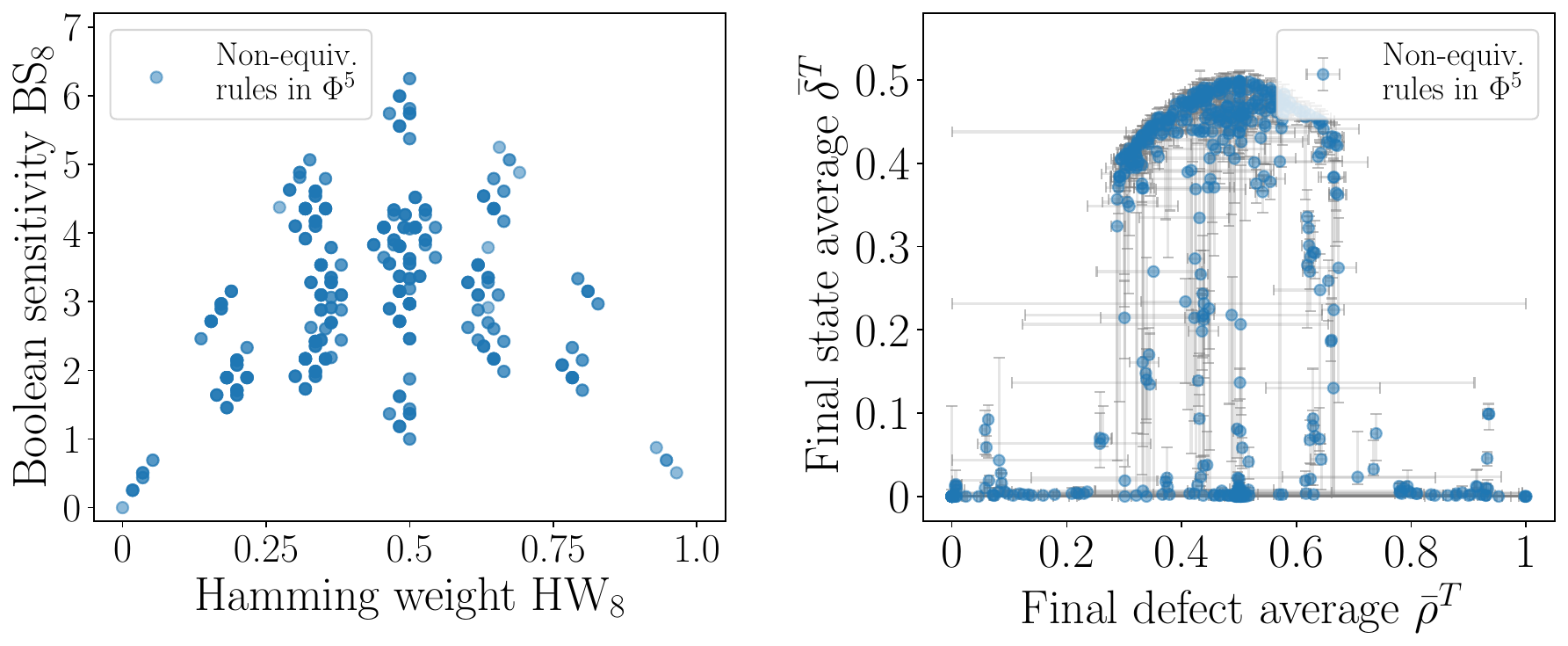}
    \caption{Scatter plot for genotype and phenotype values of all $528$ non-equivalent resolution-$5$ LLNA rules. \textit{Left}: HWs and Boolean sensitivities for degree $k=8$. Note how the $\text{HW}_8$ value limits the range of possible values for $\text{BS}_8$. \textit{Right}: Median values (circles) and IQRs (grey bars) for  to the final state and defect averages, evolved on a small-world network ($p=0.2$) from random initial configurations with $\rho^0=1/2$.}
    \label{fig:BS-vs-HW-and-state-vs-defect_median_resolution}
\end{figure}

We find a more general relationship between the  $\bar{\rho}^T$ and  $\bar{\delta}^T$ values of Life-like NA rules, which we show in Fig.~\ref{fig:BS-vs-HW-and-state-vs-defect_median_resolution} for all $528$ non-equivalent rules with resolution $r=5$. In that figure, the genotype is calculated by supposing a degree of $k=8$, while the phenotype is a result of running on the small-world networks defined in Subsec.~\ref{subsec:computational-pipeline}. Note how the shape of this relation resembles a similar relation for the genotype: the final state average limits the final defect average similarly to how the HW limits the BS value. This is a visually striking example of their mutual similarity; more quantitatively, though, the Pearson correlation coefficient between the $\text{HW}_8$ and $\bar{\rho}^T$ values is $R=0.73$, and the coefficient between the $\text{BS}_8$ and $\bar{\delta}^T$ values is $R=0.54$. Correlation values for other topologies are slightly smaller for highly regular networks, but overall quite similar. While this means that the genotype-phenotype correlation is robust, it is also somewhat surprising, considering the often strong dependence that we saw earlier of the phenotype on parameters like $p$ and $\rho^0$. This suggests that the rules with the weakest genotype-phenotype correlation are also the ones that display phase transitions. Most of all it means that while we may mobilise the genotype in order to filter out undesired regions of rule space, we must always still manually investigate individual dynamics, as some level of unpredictability remains.





\section{Finding a rule with desired properties}
\label{sec:fssp}


\subsection{The firing squad synchronisation problem}

\noindent The demonstrated correlation between genotype and phenotype can be exploited to find a particular NA that displays some desired properties. We will demonstrate this by tackling the firing squad synchronisation problem (FSSP) \cite{moore1968generalized}, which despite its simple set-up is far from trivial to solve, even for regular grids \cite{mazoyer1988overview}. In this classic problem, the goal is to iterate local updates starting from a random initial configuration, leading to a coordinated alternation between two fully homogeneous states, i.e.~where $\rho^T = 1-\rho^{T+1} = \rho^{T+2} \in \{0,1\}$ after some time $T$. Pinpointing an effective local transition rule is non-trivial because each node’s decision is confined to its \textit{local} neighbourhood, yet we require perfect \textit{global} synchrony. Applied to our binary NA, the objective therefore becomes to identify a rule $\phi$ that will generate an LLNA that effects this synchrony.\\

The FSSP is not perfectly solvable for CAs with fewer than five states \cite{mazoyer1987six}, and the generalised topologies of NAs only further complicates the mathematics \cite{nishitani1981firing}. For these reasons, the typical approach to solving this and other global computation tasks is to run a genetic algorithm (GA) \cite{mitchell1994evolving, das1995evolving}. A GA employs the blind force of competition between randomly selected rules, and outputs the best-performing competitor after a set number of generations. While this certainly leads to a high-performance solution, this top-down approach generally does not enable an intuitive understanding of \textit{why} some local interaction leads to the desired global behaviour. However, with our established prior knowledge of the behaviour associated with particular local update rules, a bottom-up approach of the FSSP becomes feasible. This essentially strikes a balance between the mathematical and the algorithmic point of view, and clearly displays the power of the genotype. Let us present a possible approach.

\subsection{The genotype of a good FSSP solver}

\noindent An efficient FSSP solver $\phi \in \Phi^r$ should possess at least two properties. First the rule should be equivalent to itself ($\phi = \phi'$), because complementing the initial configuration should have no effect on the rule's performance, as the desired final synchronous alternations are also qualitatively indistinguishable from their complement. Second, the mean-field curve must reflect the desire that the system moves towards a homogeneous state (where $\rho \in \{0,1\}$), while simultaneously alternating between the two states, so this requires that 
\begin{align*}
    \begin{cases}
        \langle\rho^{t+1}\rangle \geq 1-\rho^t & \text{if } \rho^t \leq 1/2, \\
        \langle\rho^{t+1}\rangle < 1-\rho^t &\text{else,}
    \end{cases}
\end{align*}
which imposes an unstable equilibrium at $\langle\rho^{t+1}\rangle=\rho^{t}=1/2$. In addition to these two conditions, two more genotype properties are advantageous. First, we prefer to maximise the slope $|d\langle\rho^{t+1}\rangle / d\rho^t|$ at the equilibrium, such that the system moves towards a homogeneous state fast. Second, we prefer a high BS value, which represents our desire for high volatility and chaoticity, steering away from static non-homogeneous states in which our deterministic system might get `stuck'. This is similar to how probabilistic update rules \cite{watts1999global} or the addition of noise \cite{challa2024effect} perturb the system towards stability when stochasticity is allowed.\\



\begin{figure}
    \centering
    \includegraphics[width=\linewidth]{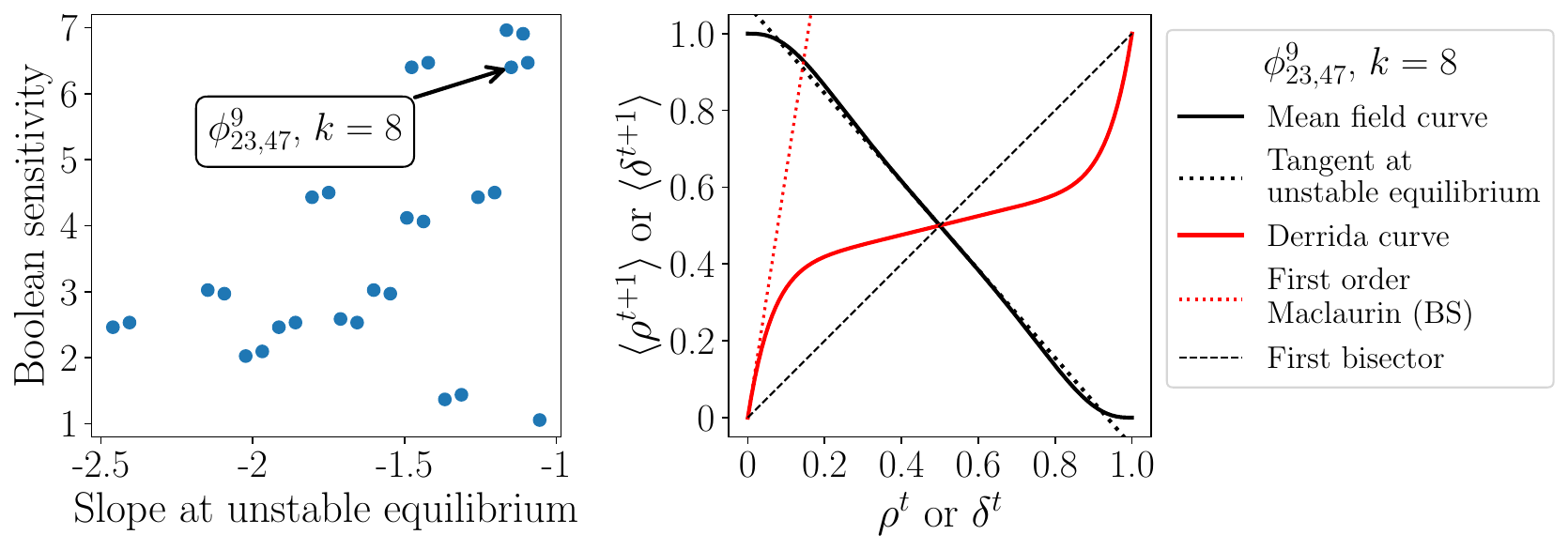}
    \caption{\textit{Left}: There are $27$ candidate rules in $\Phi^9$ whose genotype suggests that they could solve the firing squad synchronisation problem. Amongst these rules, we find a trade-off between the $\text{BS}_8$ value and the slope of the mean-field curve at $\rho^t=1/2$. \textit{Right}: The mean-field curve (black) and Derrida curve (red) of the best candidate (rule $\phi^9_{23,47}$).}
    \label{fig:fssp-27candidates-focus_on_23_47}
\end{figure}

For simplicity, we only consider a small-world network with mean degree $\langle k\rangle \approx 8$. We impose the criteria onto the mean-field curves provided by Eq.~\eqref{eq:mean_field_curve} with degree $k=8$ and resolution $r=9$. This results in $27$ candidate rules, representing a mere $0.01\%$ of all rules in $\Phi^9$. It is not a priori clear which of these candidate rules will perform best for the FSSP, because of at least three reasons. First, we observe a trade-off between the BS value and the slope of the mean-field curve, as is shown in the left-hand panel of Fig.~\ref{fig:fssp-27candidates-focus_on_23_47}. Second, the BS value is mainly relevant for configurations for which $\rho^0=1/2$, which does not necessarily reflect perturbation sensitivity in situations where one of the states dominates the configuration. Third, we have performed this preselection by supposing that the degree $k=8$, whereas we know that nodes have various degrees in a small-world network, and we have seen in Sec.~\ref{sec:genotype-parameters} that mean-field and Derrida curves may depend on the degree significantly. We therefore move our focus to the phenotype, which is now computationally feasible due to the genotype-based filter.

\subsection{Performance assessment of the candidate rules}

\noindent We inspect the phenotype of all $27$ candidate rules over the networks described in Subsec.~\ref{subsec:computational-pipeline}, now running over $T=2N=1800$ time steps like in \cite{watts1999global}. We observe that the majority of these rules does not exhibit the desired behaviour, and almost half do not generate complete synchronisation for even a single sample. One candidate rule is an excellent solver, however, namely rule $\phi^9_{23,47}$ which has a success rate of $84\%$. The mean-field and Derrida curves of $\phi_{23,47}^9$ for $k=8$ are shown on the right-hand side of Fig.~\ref{fig:fssp-27candidates-focus_on_23_47}. Surprisingly, the BS/slope trade-off is made strongly in favour of the BS value for our best-performing rule, which suggest that (for these networks) chaoticity is more important for synchronisation than mean-field evolution speed.\\

\begin{figure}
    \centering
    \includegraphics[width=.85\linewidth]{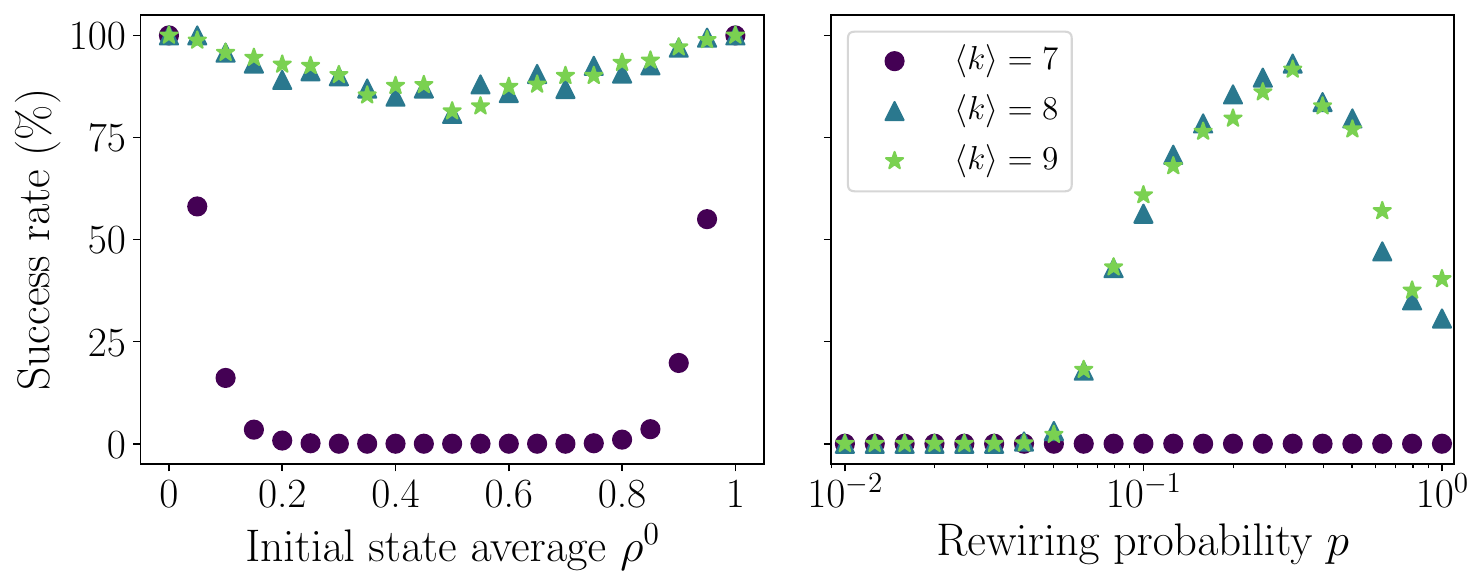}
    \caption{FSSP success rates for rule $\phi^9_{23,47}$ on various networks with $N=900$ nodes: with average degrees $\langle k\rangle = 7$ (purple circles), $8$ (teal triangles) and $9$ (green stars). The success rate for $\langle k\rangle = 7$ is always significantly lower than for the other $\langle k\rangle$ values. \textit{Left}: the state average $\rho^0$ of the initial configuration influences the success rate symmetrically around $\rho^0=1/2$. \textit{Right}: only small-world networks with an intermediate rewiring probability support emergent synchronisation.}
    \label{fig:fssp-succes_rate-degrees7_8_9-N900-T1800}
\end{figure}

A success rate of $84\%$ is impressive compared to the results of Watts's contrarian rule \cite{watts1999global}, which for $p=0.2$ networks rarely surpasses the $50\%$ mark. It is also a strong result compared to GA approaches -- which typically surpass $99\%$ but explore a much wider rule space \cite{das1995evolving} --, simply because it is \textit{so} much cheaper to calculate, and easier to interpret by means of the mean-field behaviour. Fig.~\ref{fig:fssp-succes_rate-degrees7_8_9-N900-T1800} shows that even better results are achieved when the initial state average and the topology are altered, which has also been observed in Ref.~\cite{darabos2007performance}. Unsurprisingly, a $\rho^0$ value that is close to $0$ or $1$ will result in a relatively high success rate, simply because part of the work towards configurational homogeneity has already been done. More interesting is the dependence on the average degree and the rewiring parameter, clearly indicating the sudden change in behaviour when moving from $\langle k \rangle =7$ to $8$, and -- surprisingly -- \textit{decreasing} in performance when $p \to 1$. We find a maximal success rate of $93\%$ for small-world networks with $\langle k\rangle=8$ and $p \approx 0.32$. This plot demonstrates once more that the LLNA's phenotype can depend strongly on topological parameters that are not embedded in the genotype of the rule itself, even allowing for phase transitions. It therefore remains very important to verify the expected behaviour in practice.\\


As a concluding remark, we highlight that the symmetry operations discussed in Subsec.~\ref{subsec:relationships-limits-symmetries} allow us to use the genotype of FSSP solvers for similar problems such as the density classification task \cite{mitchell1994evolving}. After all, if we want the system to move towards a \textit{static} homogeneous state, we want to maintain the Derrida curve that was pertinent for the FSSP, and vertically flip the mean-field curve, such that $\langle\rho^{t+1}\rangle = 0$ and $\langle\rho^{t+1}\rangle = 1$ are stable equilibria. This implies that we perform the map $(B, S) \mapsto (B^\text{C}, S^\text{C})$. While we observe that the resulting rules often lead to wrong predictions, we will see in future work that the mere property of steering towards homogeneous end states can be employed as an invaluable tool in the analysis of complex networks themselves. These are just some of the many possibilities enabled by serious consideration of the correlations between the LLNA's genotype and phenotype.

\section{Conclusion and future work}
\label{sec:discussion}

\noindent In this article we covered a thorough theoretical basis for outer-totalistic NAs. We did so in four phases.
First, we introduced  an internally consistent topological generalisation of LLCAs, by making sure that the model reduces to the classical CA case for lattice networks, and by establishing complementation symmetry in the definition of the local update rule. This clearly establishes the model in the landscape of discrete systems by highlighting the models it is related to (e.g.~CAs) and the models it encompasses (e.g.~threshold models).
Second, we investigated the model's genotype by defining the HW and the BS for outer-totalistic update rules. In particular, we showed how a mean-field approach can help generate a mean-field curve and a Derrida curve for any and every update rule and topology, which is a powerful analytical tool in the study of these models.
Third, we mapped out the phenotype of this model, by computing the actual state evolution of the model for various initial configurations and topologies. This clearly demonstrated that the same local update rules can generate vastly different behaviour for often merely slightly different topological and configurational changes.
In a fourth phase, we highlighted how the genotype and phenotype are intimately connected, particularly how the HW is correlated to the final state average, and how the BS is correlated to the final defect average. Although the inherent complexity of these kinds of systems restricts a clean one-to-one relationship between genotype and phenotype, we demonstrated the utility of our results by engaging in a bottom-up approach to solving the firing squad synchronisation problem, with convincing success rates of over $90\%$.\\

Some natural next steps present themselves. Purely in terms of the dynamical model itself, many further extensions are also relatively straightforward. As we showed in \cite{rollier2025comprehensive}, a generalisation to probabilistic, continuous, asynchronous, and/or non-uniform NAs is easily made and often required for practical applications and meaningful investigations of complexity. Here we intentionally investigated the simplest model extension, only the essential dynamical metrics, and only varied over a single topological parameter (the Watts-Strogatz rewiring probability). That condenses into arguably the main general contribution of this article: facilitating the analysis of this broad class of discrete dynamical models by means of a set of essential mathematical tools.\\

Clearly, there are many more particular relations to uncover. We see at least three interesting research possibilities. First, a more detailed analysis of the time series associated with different rules and topologies, for example by means of detrended fluctuation analysis aimed at identifying long-range temporal correlations \cite{peng1994mosaic}. This would help mitigate the problem that the final medians $\bar{\rho}^T$ and $\bar{\delta}^T$ ignore valuable information of the model's rich dynamics. Second, we may investigate alternative metrics for the defect spread, the most interesting option being the Lyapunov spectrum in the tangent space interpretation of defect propagation \cite{vispoel2024damage}, which should give a more fine-grained quantification of chaoticity. Third, the introduction of seeds and defects can be explicitly related to node properties such as the degree or the Page rank \cite{gleich2015pagerank} for various network types, including networks with high community structure where defect propagation is not at all obvious \cite{karrer2011stochastic}. This would disentangle the damage spread shown in this article, where defects were introduced randomly, rather than based on structural information. As a result, this may further a more detailed understanding of the influence of individual nodes on the network dynamics, i.e.~the influence of the parts on the whole. That exciting abstract investigation into targeted impact is the focus of our upcoming research paper.

\section*{CRediT authorship contribution statement}

\noindent \textbf{Michiel Rollier:} Conceptualization, Investigation, Methodology, Software, Visualization, Writing -- Original Draft, Writing -- Review \& Editing. \textbf{Lucas Caldeira de Oliveira:} Software, Writing -- Review \& Editing. \textbf{Odemir M.~Bruno:} Funding acquisition, Supervision. \textbf{Jan M.~Baetens:} Conceptualization, Funding acquisition, Supervision, Writing -- Review \& Editing.

\section*{Declaration of competing interest}

\noindent The authors declare that we have no known competing financial interests or personal relationships that could have appeared to influence the work reported in this paper.

\section*{Data availability}

\noindent No non-synthetic data was used for the research described in this article.

\section*{Acknowledgements}

\noindent This work has been partially supported by the FWO grant with project title ``An analysis of network automata as models for biological and natural processes'' [3G0G0122]; by the FWO travel grant with file name V412625N; by the FWO congress participation grant K105625N; by the FAPESP grant \#2024/02727-0; by the CAPES grant \#88887.841805/2023-00. The authors wish to thank Gisele H.~B.~Miranda and Bernard De Baets for their invaluable contribution to the mathematical foundations of LLNAs.

\ \\
\hrule





\bibliographystyle{elsarticle-num-names} 

\begin{flushleft}
\setstretch{1.0}
\footnotesize
\begin{multicols}{2}
\bibliography{outer-totalistic-network-automata}
\end{multicols}
\end{flushleft}

\end{document}